# Consistency and validity of interdisciplinarity measures

Qi Wang [1, 2, 3, *]   Jesper Wiborg Schneider [1, #]

*qiwang@kth.se; jws@ps.au.dk*

[1] The Danish Centre for Studies in Research and Research Policy (CFA), Aarhus University, Denmark
[2] KTH Library, KTH-Royal Institute of Technology, Sweden
[3] Department of Philosophy and History, KTH-Royal Institute of Technology, Sweden
*ORCID: 0000-0001-7817-5327
#ORCID: 0000-0001-5556-0919

**Abstract**

Measuring interdisciplinarity is a pertinent but challenging issue in quantitative studies of science. There seems to be a consensus in the literature that the concept of interdisciplinarity is multifaceted and ambiguous. Unsurprisingly, various different measures of interdisciplinarity have been proposed. However, few studies have thoroughly examined the validity and relations between these measures. In this study, we present a systematic review of these interdisciplinarity measures and explore their inherent relations. We examine these measures in relation to the Web of Science journal subject categories. Our results corroborate recent claims that the current measurements of interdisciplinarity in science studies are both confusing and unsatisfying. We find surprisingly deviant results when comparing measures that supposedly should capture similar features or dimensions of the concept of interdisciplinarity. We therefore argue that the current measurements of interdisciplinarity should be interpreted with much caution in science and evaluation studies, or in relation to science policies. We also question the validity of current measures and argue that we do not need more of the same, but rather something very different in order to be able to measure the multidimensional and complex construct of interdisciplinarity.

## 1. Introduction

Studies that examine "interdisciplinarity" quantitatively tend to bemoan the measurement situation. Criticisms are rife and there is no consensus in relation to the definition and operationalization of interdisciplinary research (IDR) (e.g. Rafols et al., 2012; Wagner et al., 2011). As a consequence, numerous indicators or metrics purport to measure the concept or aspects of it.

The concept of interdisciplinarity is tied to notions of academic disciplines. Essentially the concept is often envisioned as a synthesis of theories or methodological activities from different disciplines resulting in an emergent interdisciplinary activity. However, there is considerable ambiguity with the discipline concept itself, its delineation and empirical manifestations (Sugimoto & Weingart, 2015). Historically, disciplines have been linked to the organization of teaching at universities. However, there is more to disciplines than the fact that something is a subject taught in an academic setting. The concept has evolved to become a more general term encompassing the organization of learning, but also the systematic production of new knowledge (Abbott, 2001). Focusing on knowledge production, Sugimoto and Weingart (2015) suggest that academic disciplines can be examined empirically from three perspectives, what they term "publications", "people" and "ideas". In essence, all three perspectives rely on data from (journal) publications in multidisciplinary bibliographic databases. What Sugimoto and Weingart (2015) refer to as "publications" are disciplinary delineations based on indexing and classification of publications and/or their parent journals. "People" uses authors, mentors, and affiliations as delineations, whereas "ideas" refer to cognitive attributes such as language use, topics, and methodology.

Most measures of interdisciplinarity are rooted in scientometric operationalizations of disciplinary structures, mainly based on the "publication" perspective suggested by Sugimoto and Weingart (2015). The scientometric approaches utilize the vast quantities of publications in bibliographic databases to structure the literature according to established categories or by applying various different indexing techniques. Depending on the purpose, such structures are then perceived as "disciplinary structures". However, with no conceptual or operational consensus and plenty of attributes and researcher degrees of freedom, it is no surprise that IDR based on scientometric techniques has been scrutinized and interpreted in many different ways. Numerous indicators, measures or metrics have been proposed for measuring interdisciplinarity, but only a few studies have actually examined the relation between such measures, their validity, and consistency. While Rafols and Meyer (2007) initially concluded that interdisciplinarity measures based on publications and their citation relations can provide a comparatively accurate description of cross-boundary knowledge creation, Leydesdorff and Rafols (2011, p. 98) later concluded that "different indicators may capture different understandings of such a multi-faceted concept as interdisciplinarity". Recently, a report by Digital Science (2016, p. 2) concluded that the choice of datasets and methodologies produces "inconsistent and sometimes contradictory" results.

The aim of the present study is to further examine the relations between a number of seemingly similar interdisciplinarity measures. We examine empirically to what extent findings based upon them are consistent, in order to be able to shed more light on their validity and potential use for science policy. We limit our empirical review to proposed interdisciplinarity measures based on the "publication" perspective, i.e. publications and citation relations.

The article is structured as follows: In Section 2 we briefly introduce the data used for the empirical analyses. In Section 3 we summarize definitions and measures of interdisciplinarity examined in the study. Subsequently, we present the results in Section 4. Discussion and conclusions follow in Sections 5 and 6.

## 2. Unit of analysis

In the present study, we examine a number of known interdisciplinarity measures empirically by comparing their outcomes when applied to the Web of Science (WoS) journal subject categories (SCs). As data, we use all publications of the document type *article* published in 2010 from the in-house WoS database at the Centre for Science and Technology Studies (CWTS) at Leiden University. As the validity and effectiveness of using a journal citation database is questionable for research fields where journals are not the main scientific communication medium, we exclude

SCs from the Arts & Humanities Citation Index, resulting in a total of 224 WoS SCs included in the analysis.

Classification systems are essential when quantifying interdisciplinarity using bibliometric methods. The groupings are presumed to represent "disciplines" and their potential mutual relations become the determining factor when measuring "interdisciplinarity". Numerous classification systems are available. In this study, we use the WoS journal classification system as disciplines or categories. The WoS classification is by no means a "ground truth", on the contrary, it is arbitrary in its details (Wang & Waltman, 2016). Indeed, no classification system can be seen as the "truth", different systems may serve different purposes and, in that sense, the choice of a system should be seen in relation to the purpose of a study. For instance, the lower level of the Organisation for Economic Co-operation and Development (OECD) classification system has around 40 categories, whereas the WoS classification system consists of 250 SCs. Following the above logic, a unit of analysis may show a higher degree of interdisciplinarity when using the WoS system compared to the OECD system simply due to their different granularities. We use the WoS classification system because it is generally the most frequently used system in scientometric studies, including interdisciplinarity research based on bibliometric methods.

Before we introduce measures of interdisciplinarity, two issues need to be stressed. First, references from, and citations to, scientific publications can both be used to operationalize interdisciplinarity. According to Levitt et al. (2011, p. 1121), "there does not seem to be clear evidence that one is preferable to the other", however, others stress that citations and references have different implications (e.g. Porter & Chubin, 1985). We believe that references given in a publication are a reflection of the "knowledge base" upon which the work is built. Compared to citations, reference lists in publications better reflect the potential integration of knowledge from different research fields, a focal issue in relation to measuring interdisciplinarity. Consequently, we focus on the use of references in this study.

Second, most interdisciplinarity measures can be applied at the level of individual publications but also at higher levels of aggregation. For example, for a set of publications belonging to a unit of analysis, one can apply the Rao-Stirling index (RS) to measure the degree of interdisciplinarity for each of the publications individually and subsequently compute the mean or median and use it as the degree of interdisciplinarity for the unit. Alternatively, one could instead view all publications from the unit of analysis as one entity under which all references are subsumed. Then, the proportion of these references over different WoS SCs would be used as the input for the RS index. We use the latter approach in this study. However, the RS index will also be used to demonstrate differences in degrees of interdisciplinarity when measured at different levels. We elaborate on this in the following section.

## 3. Overview of interdisciplinarity measures

In this section, we first review and discuss the definitions of IDR used in bibliometric studies. Subsequently, we summarize the proposed interdisciplinarity measures. We briefly discuss the reviewed measures at the end of this section.

*3.1. Related work on definitions and operationalizations of IDR*

We have identified a seed set of publications studying interdisciplinarity in bibliometric studies based on our prior knowledge. The set was expanded by including pertinent publications according to the references given in the seed publications. The definitions of IDR were extracted from 15 publications that are outlined chronologically in Table A1 in the Appendix.

While the overall concept of IDR is seen by some as ambiguous or uncertain (e.g. Rafols et al., 2012; Wagner et al., 2011), the impression one gets from reading the definitions summarized in Table A1 is actually one of small nuances between them. They are quite similar in their conceptualization seeing IDR as a kind of knowledge integration. It seems that especially two key attributes are frequently emphasized, "diversity", which describes the differences in the bodies of knowledge that are integrated and "coherence", which describes the intensities of the relations between these bodies of knowledge. Diversity is presumed to reflect to what extent a body of knowledge or a unit of analysis comprises knowledge rooted in two or more different research fields. It seems to be the most important attribute of IDR, as almost all studies reviewed discuss it.

According to Rafols et al. (2012), the "integration" of research is perceived as the process of establishing connections between cognitively distant or separate bodies of knowledge. Some argue that integration (i.e. coherence) is a necessary complement to diversity in order to identify IDR (e.g. Rafols & Meyer, 2010). While a high degree of diversity implies that a body of knowledge draws upon knowledge from several fields, it does not indicate to which extent, if at all, such knowledge is mutually integrated. In this view, IDR is seen as a combination of a high degree of diversity and coherence (i.e. knowledge integration) (Rafols & Meyer, 2010; Rafols et al., 2012).

One can argue that the definition and operationalization of IDR by Leydesdorff (2007) differs from the other listed studies, as the degree of interdisciplinarity depends on a unit's position in a citation network. Rafols et al. (2012) use the notion of "intermediation" to refer to this network perspective of interdisciplinarity. However, in our view, the operationalization of intermediation is essentially related to diversity. For example, a journal's degree of interdisciplinarity is based upon its relations to adjacent journals in a citation network. More citation links for a journal implies a higher degree of interdisciplinarity. In that sense, we think that intermediation depicts diversity externally to a body of knowledge, and not internally, as initially intended with the diversity definition state above, where the focus is upon knowledge heterogeneity within a body of knowledge.

Finally, some terminological inconsistencies exist. Many near-synonyms are used to describe attributes of interdisciplinarity or its opposite features, for instance, "specialization", "concentration", "unevenness", "information richness", "information abundance" etc. As they are poorly delineated and most likely redundant in relation to the above-mentioned main features, we exclude these variant terms from our analyses.

In the following subsections, we classify the interdisciplinarity measures proposed from previous studies into four groups based on the resemblance of their strategies. This grouping is subjective and serves to improve the reading flow of this section. Other groupings can be conceived of. Furthermore, given the limited space, it is not possible to make exhaustive descriptions of each measure. For more detailed descriptions we refer to the previous studies.

*3.2. Group 1: Interdisciplinarity measures depending on a multi-classification system*

The first group of measures are characterized by depending to a large extent on the WoS journal classification system for their calculation. As the WoS SC indexing is not exclusive, journals can be assigned to multiple categories. The multiple indexing strategy is utilized by some to measure interdisciplinarity. The relevant measures are briefly introduced below.

- *Percentage of multi-assigned journals* (p_multi). Let $i$ denote a WoS SC. P_multi is the percentage of journals in $i$ that have been assigned to more than one SC (Morillo et a., 2001; 2003).
- *Percentage of journals outside the area* (p_outside). The percentage of journals in $i$ that have been assigned to more than one research area. Research areas are higher aggregation levels consisting of several SCs. Notice, such levels are not part of the WoS system and need to be constructed (Morillo et a., 2001; 2003). This study applies the CWTS high aggregation of SCs as research areas.
- *Percentage of references outside the category* (pro). The percentage of journal references that publications in SC $i$ cited outside $i$ (Morillo et a., 2001; Porter & Chubin, 1985).
- *Diversity of links* (d_links). The number of distinct journal pairs generated from journals belonging to SC $i$, in which the pair of journals should belong to different SCs. To reduce size effects, it is normalized by the total number of journals in SC $i$ (Morillo et al., 2003).
- *Pratt index*. The Pratt index was initially proposed to measure concentration that allows comparison of SCs and journals in different research fields (Pratt, 1977). Accordingly, the higher the proportion of references from different SCs that publications in SC $i$ cited, the more interdisciplinary SC $i$ is considered. Since this index has a negative relation with interdisciplinarity, we use 1-Pratt instead.
- *Specialization index* (Spec). Porter and colleagues suggested that the exploration of specialization could provide insight into IDR (Porter et al., 2007). They proposed the specialization index to measure the spread of references that publications in SC $i$ cited over all other SCs. The specialization index is very similar to the Pratt index, but perhaps more intuitive. Like the Pratt index, the specialization index is also inversely related to interdisciplinarity, and hence we use 1-Spec instead.

*3.3. Group 2: Interdisciplinarity measures borrowed from other fields*

Indices originating in other fields such as economics and biology have also been suggested as measures of interdisciplinarity. These indices were originally constructed to measure very different constructs such as biodiversity, income inequality, and information uncertainty to name three well-known examples. They are briefly summarized below.

- *Simpson diversity index*. Simpson diversity index measures the probability that two entities randomly sampled from a population will not belong to the same category. (Simpson, 1949; see also Zhang et al., 2016).
- *Shannon entropy*. Shannon entropy was proposed to measure "information uncertainty" (Shannon, 1948; 2001). Some argue that "information uncertainty" is linked to the concept of diversity, since entropy can quantify the distribution of references over SCs. This is to some extent similar to the design of 1-Pratt and 1-Spec. For example, if publications in a category only cited other publications in this category, the diversity of the reference distribution would be maximal and the uncertainty would be minimal (Leydesdorff & Rafols, 2011).
- *Brillouin diversity index.* Brillouin index is a modification of Shannon entropy, and also aims to measure the uncertainty of information (Brillouin, 1956). Steele and Stier (2000) argued that "the Brillouin index is a proper indicator of interdisciplinarity since it considers the number of observations and the distribution of observations among categories" (Huang & Chang, 2012, p. 793).
- *Gini coefficient.* The Gini coefficient was proposed as a measure of income inequality. When assessing interdisciplinarity, it considers the distribution of references over SCs for a group of publications (e.g. Leydesdorff & Rafols, 2011; Wang et al., 2015). This is also to some extent similar to the design of 1-Pratt and 1-Spec. It should be noted that the Gini coefficient has a negative relation with interdisciplinarity, and thus we use 1-Gini.

*3.4. Group 3: Interdisciplinarity measures that consider the similarity of research fields*

The measures introduced so far focus on the overlap of publications or references in SCs and/or the distribution of references over SCs. Such measures are criticized for not considering the similarity of SCs. Consequently, interdisciplinarity measures that include a similarity index have been proposed, including the three measures below.

- *RS index.* The RS index has been widely used to measure diversity and more generally, interdisciplinarity (e.g. Porter et al., 2007; Porter & Rafols, 2009; Wang et al., 2015). It is assumed that the index incorporates essential attributes of diversity i.e. variety, balance, and similarity (Rafols & Meyer, 2010). We will not discuss the attributes of diversity further in this study. However, since the construction of similarity matrices and the influence it has upon the RS is often neglected in interdisciplinarity studies, we briefly discuss this issue at the end of this section.
- *Hill-type measure.* Recently, Zhang et al. (2016) claimed that the RS index produces results with low discriminative power. They stated, "[a]s seen in the study by Zhou, Rousseau, Yang, Yue, and Yang (2012), the Rao-Stirling measure showed only low discriminatory power since values, at least in their work, sometimes differ only by the third decimal" (p. 1257). The low discriminative power among the interdisciplinarity values may therefore cause problems in practical applications (see also Zhou et al. 2012). To overcome the presumed limitations of RS, the Hill-type measure was proposed (Hill, 1973; Leinster & Cobbold, 2012; Zhang et al., 2016).
- *Coherence measure.* This measure emphasizes the knowledge integration between different research fields. It was used, for instance, by Soos and Kampis (2012) and Wang (2016). We name it a coherence measure. In the present study, the coherence of SC *i* was examined based on the references of publications belonging to *i*. The more intensive the citation links of the references belonging to different SCs, the higher the knowledge integration between research fields and the larger the degree of interdisciplinarity for SC *i* is considered. Note, we do not distinguish between the directions of citation links of the references here. This is unnecessary as we are only interested in the degree of integration of references from different SCs. Finally, some claim that this measure actually combines diversity and coherence, and hence argue that it can be used exclusively for measuring interdisciplinarity (Soos & Kampis, 2012; Wang, 2016).

*3.5. Group 4: Interdisciplinarity measures that rely on networks*

The measures we introduced above can be seen as "within-based" measures because they rely upon publications and reference relations within a set of publications. The global SC network, i.e. the bibliometric relations between the SC under investigation and the external SCs, are rarely included in the discussion of interdisciplinarity. As discussed in Section 2, some argue that the location of a category in a global network can indicate its degree of interdisciplinarity (Leydesdorff, 2007; Rafols, et al., 2012). More specifically, it is assumed that when a group of publications are located in an intermediate position in a network, it is an indication of IDR (Rafols, et al., 2012). Such types of interdisciplinarity measures are listed below.

- *Betweenness-centrality* (BC). Leydesdorff (2007) proposed to use the BC index (Freeman, 1977) to measure the degree of interdisciplinarity of journals. Betweenness measures the degree of centrality that a node (entity) is located on the shortest path between two other nodes in a network (Freeman, 1997). Furthermore, if a journal or a SC is at the intermediate position between other journals or SCs, then the journal or the SC can be considered as interdisciplinary, and its publications function as a communication channel for other journals or SCs (Leydesdorff, 2007; Silva et al., 2013).
- *Cluster coefficient* (CC). This measure was introduced by Rafols et al. (2012). For a given SC, it first identifies the proportion of observed references between this category and other SCs over the expected maximum number of references. The proportion is then weighted by the percentage of publications that this SC has over the total number of publications. The cluster coefficient of this category is the sum of these weighted proportions to other different SCs.
- *Average similarity* (AS). It was also introduced by Rafols et al. (2012). For a given SC, it simply measures the average similarity of this SC to all other SCs, and weights by the percentage of publications that this category has over the total number of publications. The average similarity of this category is the sum of these weighted similarities.

*3.6. Summary of interdisciplinarity measures*

We have outlined 16 interdisciplinarity measures that use publications and reference relations. Table 1 lists the notation used in this study. Table 2 presents the measures and their formula. Despite our attempt to cover all interdisciplinarity measures in bibliometric studies, not all of these are included in the present work for several reasons. For instance, Mugabushaka et al. (2016) examined the use of different threshold values for the parameter in the Hill-type measure and concluded that the differences are in fact very small. Hence, we will not do a further test on other Hill-type measures in the present work.

Table 1. Notation in this study

| Notation | Description |
| --- | --- |
| $a_i$ | The number of publications in category $i$ |
| $c_{ik}$ | The number of references from publications in category $i$ that cited publications in category $k$ |
| $r_{jik}$ | The number of the shortest paths from categories $j$ to $k$ that pass through category $i$ |
| $r_{jk}$ | The number of shortest paths between categories $j$ and $k$ |
| $p_{ik}$ | The proportion of references from publications in category $i$ that cited publications in category $k$, it can be expressed as $p_{ik} = c_{ik}/\sum_j c_{ij}$ |
| $q_{ik}$ | The proportion of references from a publication in category $i$ that cited publications in category $k$, $q_{ik}$ is measured at the individual publication level |
| $P_i$ | The proportion of publications in category $i$ over the total number of publications of all SCs, it can be expressed as $P_i = a_i/\sum_j a_j$ |
| $s_{ij}$ | The similarity between categories $i$ and $j$ |
| $c_{jk}^i$ | For publications in SC $i$, the number of citation links between their cited references in categories $j$ and $k$ |
| $n$ | The number of SCs that the references of the publications in SC $i$ belong to |

Table 2. Interdisciplinarity measures reviewed in this study

| Measure | Formula |
|---|---|
| p_multi | The percentage of multi-assigned journals |
| p_outside | The percentage of journals that are classified in more than one research area |
| pro | $\sum_{k \neq i} c_{ik} / \sum_j c_{ij}$ |
| d_links | The number of links between different SCs established by journals in a given category |
| Pratt index | $\frac{2^{(n+1)}/2 - \sum_k g p_{ik}}{n-1}$, where $g$ is the index obtained by ranking $p_{ik}$ in decreasing order |
| Spec | $\sum_k c_{ik}^2 / (\sum_k c_{ik})^2$ |
| Simpson index | $1 - \sum_k p_{ik}^2$ |
| Shannon entropy | $-\sum_k p_{ik} \ln p_{ik}$ |
| Brillouin index | $(\log(\sum_k c_{ik})! - \sum_k \log c_{ik}!)/\sum_k c_{ik}$ |
| Gini coefficient | $\frac{\sum_k (2h - n - 1) c_{ik}}{n \sum_k c_{ik}}$, where $h$ is the index attained by sorting SCs according to $c_{ik}$ in increasing order |
| RS | $\sum_{j,k}(1 - s_{jk}) p_{ij} p_{ik}$ |
| Hill-type measure | $1/\sum_{j,k} s_{jk} p_{ij} p_{ik}$ |
| Coherence | $\sum_{j,k} c_{jk}^i (1 - s_{jk})$ |
| BC[1] | $\sum_{j,k} \frac{r_{jik}}{r_{jk}}$ |
| CC | $\sum_j P_j \frac{c_{ij}}{a_i a_j}$ |

---

[1] Note, Leydesdorff (2007) used a symmetrical cosine matrix instead of a citation matrix to measure the interdisciplinarity of journals using BC. He claimed that the size of journals was "controlled" in this way. However, some researchers still use a citation matrix to measure BC, for instance Silva et al. (2013). In our view, it is more intuitive to use a citation matrix as input for BC as it better demonstrates the concept of intermediation. Hence, we use a citation matrix in the present study. The shortest path indicates the path with the lowest total edge weight. However, in our case, results obtained by BC can be scaled with size, i.e. if SC $i$ has a large number of publications, it is likely to have a high degree of interdisciplinarity when BC is used.

BC was calculated using the R package SNA: Tools for social network analysis. The input citation matrix was generated using SQL from the in-house WoS database at CWTS. Other measures were calculated using SQL or a combination of SQL and R.

| AS | $\sum_i P_i(\frac{1}{N}\sum_j s_{ij})$, where $N$ is the number of all other SCs |

*3.7. Further discussion on similarity and dissimilarity matrices*

It is necessary to elaborate on different approaches to generate similarity and dissimilarity matrices. While the so-called Salton's cosine similarity index (Salton & McGill, 1983) is frequently applied in bibliometric analyses, it actually has several different variants, and consequently, different solutions and results can be expected (Schneider & Borlund, 2007a; 2007b). Here, we first discuss two variants of the cosine formula. Suppose we aim to construct a symmetric similarity matrix of SCs [$s_{ij}$] based on their mutual citation relations. The first step is to construct a transaction matrix of citation relations between SCs [$c_{ij}$]. Note that [$c_{ij}$] is an asymmetric matrix with self-citations in the diagonal, whereas the similarity matrix, [$s_{ij}$], is a symmetric matrix.

One application of Salton's cosine index can be illustrated as follows. Two SCs are considered to be strongly related if they commonly cite the same SCs, i.e. their vector profiles are similar. Hence, the similarity of two SCs *i* and *j* is given by

$$S_{C(i,j)} = \frac{\sum_k c_{ik} c_{jk}}{\sqrt{(\sum_k c_{ik}^2)(\sum_k c_{jk}^2)}} \quad (1)$$

This vector application is closely aligned to the original application suggested by Salton and McGill (1983) in relation to the Vector Space Model used in information retrieval[2]. This approach is used in some bibliometric studies e.g., Leydesdorff and Rafols (2011). Another application of the cosine index is based on binary or scalar values (also known as the Ochiai index):

$$S_{O(i,j)} = \frac{c_{ij} + c_{ji}}{\sqrt{(\sum_k c_{ik} + \sum_k c_{ki})(\sum_k c_{jk} + \sum_k c_{kj})}}, i \neq j \quad (2)$$

Here, $c_{ij} + c_{ji}$ is equal to the total number of citations between SCs *i* and *j*. Note that $S_{O(i,i)}$ is set to 1. This way of calculating similarity was used in the study of Zhang et al. (2016).

In addition, several strategies can transform a similarity matrix into a dissimilarity matrix. The frequently applied solution is to use $1 - [s_{ij}]$ to obtain a dissimilarity matrix. There are also studies using $1/[s_{ij}]$ (Jensen & Lutkouskaya, 2014). In this study, RS is used as an example to demonstrate the potential empirical differences in interdisciplinarity at different levels of aggregation that may result from choosing different dissimilarity matrices. Several combinations of RS are summarized in Table 3, in which RS_P indicates the average RS interdisciplinarity of publications in a SC. Instead, RS_G views publications in a SC as one entity and uses the proportion of all its references over different SCs as the input for the RS index.

---

[2] Salton and McGill's approach, however, is more in line with traditional matrix algebra where an asymmetric data matrix of publications and terms, $n \times m$, is transformed into a symmetric similarity matrix, $n \times n$.

Table 3. RS Combinations

| RS index | Formula |
|---|---|
| RS_P[1-$S_C$] | $(1/a_i) \sum_i (\sum_{j,k} (1 - S_{C(j,k)}) q_{ij} q_{ik})$ |
| RS_P[1-$S_O$] | $(1/a_i) \sum_i (\sum_{j,k} (1 - S_{o(j,k)}) q_{ij} q_{ik})$ |
| RS_P[1/$S_C$] | $(1/a_i) \sum_i (\sum_{j,k} (1/S_{C(j,k)}) q_{ij} q_{ik})$ |
| RS_P[1/$S_O$] | $(1/a_i) \sum_i (\sum_{j,k} (1/S_{o(j,k)}) q_{ij} q_{ik})$ |
| RS_G[1-$S_C$] | $\sum_{j,k} (1 - S_{C(j,k)}) p_{ij} p_{ik}$ |
| RS_G[1-$S_O$] | $\sum_{j,k} (1 - S_{O(j,k)}) p_{ij} p_{ik}$ |
| RS_G[1/$S_C$] | $\sum_{j,k} (1/S_{C(j,k)}) p_{ij} p_{ik}$ |
| RS_G[1/$S_O$] | $\sum_{j,k} (1/S_{O(j,k)}) p_{ij} p_{ik}$ |

## 4. Results

First, we present the results regarding the relations between the interdisciplinarity measures examined; then we outline their distributions over the WoS SCs. Finally, we present an in-depth analysis of five selected WoS SCs.

*4. 1. Relations of interdisciplinarity measures*

First, we examine the consistency of the four dissimilarity measures when applied to the 224 WoS SCs. This is shown in Table 4 below using Pearson's correlation coefficients. We find inconsistencies in the dissimilarity matrices that use different versions of the cosine formulas. Interestingly, we also find inconsistencies among the ones using the same cosine formulas. As dissimilarity is an essential element for interdisciplinarity measures such as RS, variations in these matrices obviously influence the resulting interdisciplinarity values (Schneider & Borlund, 2007a; 2007b).

Table 4. Pearson's correlation coefficients for four dissimilarity measures

|  | $1 - S_c$ | $1/S_c$ | $1 - S_o$ | $1/S_o$ |
|---|---|---|---|---|
| $1 - S_c$ | 1 | | | |
| $1/S_c$ | 0.35 | 1 | | |
| $1 - S_o$ | 0.54 | 0.13 | 1 | |
| $1/S_o$ | 0.12 | 0.3 | 0.04 | 1 |

Fig. 1 shows the distributions of pairs of SCs over dissimilarity values. All distributions are skewed. The matrices based on $S_o$ as the input are extremely skewed. For instance, dissimilarity values yielded by 1-$S_o$ are largely between 0.95 and 1, which in this case implies that RS would be very close to the Simpson index.

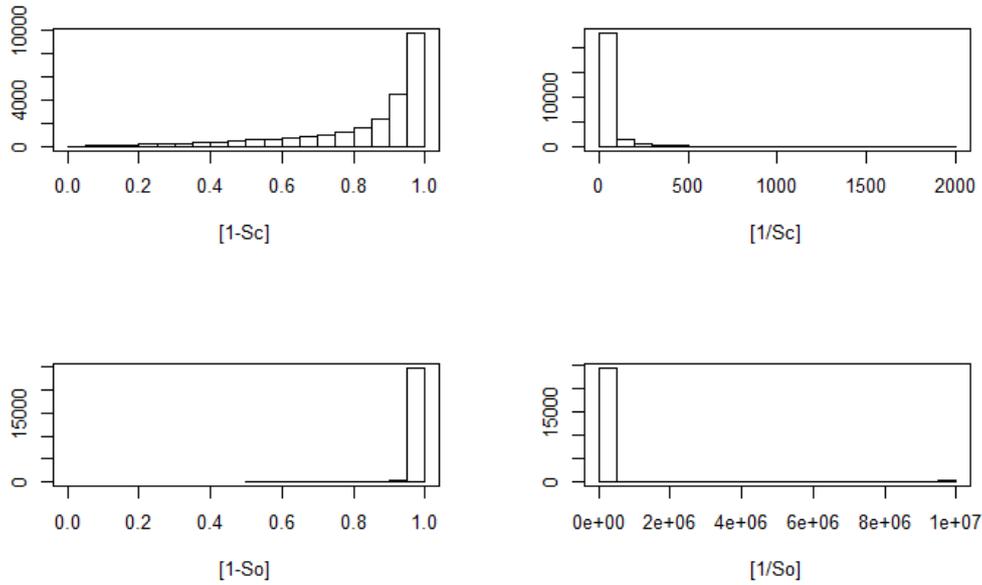

Fig. 1. Distributions of pairs of WoS SCs over dissimilarity values

As all reviewed measures are supposed to capture interdisciplinarity, we expected that their mutual correlations would be high. We use Pearson's and Spearman's correlation coefficients to examine the correlations between these measures when applied to the 224 WoS SCs. The results we obtained by using both methods are in a good agreement. Therefore, we will mainly explain the Pearson's results shown in Table 5. The Spearman's results can be found in Table A2 in the Appendix. First, we examine the measures in the first two groups, which do not rely on dissimilarity indices or global networks. According to the correlation coefficients, these measures can be roughly put into two clusters: 1) p_multi, p_outside, d_links, 1-Spec which are moderately correlated among each other; and 2) pro, 1-Pratt, Simpson measure, Shannon entropy, Brillouin index, and 1-Gini which are likewise moderately correlated among each other. As discussed above, the Pratt index is to some extent similar to the Spec measure and the Gini coefficient. Thus, a high positive correlation among them should be expected. The empirical results differ from our expectations, showing that 1-Pratt has weak correlations with 1-Spec or 1-Gini. However, measures of Shannon and Brillouin are perfectly linearly correlated.

Next, we examine the different combinations of RS. On the one hand, measures using the same dissimilarity indices tend to have higher correlations also when calculated at different levels of aggregation (i.e. individual vs. aggregated). On the other hand, measures based on different dissimilarity matrices show inconsistent results even at the same level of analysis. For instance, measures $RS\_P[1-S_c]$ and $RS\_P[1-S_o]$ both take the average RS value of individual publications as the degree of interdisciplinarity for the SCs. However, due to the differences in the dissimilarity matrices, their mutual correlation coefficient is only 0.18. As expected, different dissimilarity matrices influence RS outcomes significantly. This is in line with the conclusion of Leydesdorff and Rafols (2011). Furthermore, it was also expected that $RS\_R[1-S_o]$ and $RS\_G[1-S_o]$ show correlations with the Simpson diversity measure because $[1-S_o]$ is highly left-skewed.

The Hill-type and coherence measures also take the dissimilarity of SCs into consideration. Since the two measures apply the dissimilarity matrix $1-S_c$, they are strongly linearly correlated with the other measures using the same matrix. Note that interdisciplinarity measures relying on dissimilarities are somehow correlated with each other (except the ones using $1-S_o$ as a dissimilarity matrix), but poorly correlated with most other measures.

It is difficult to interpret the correlations between the interdisciplinarity measures in the fourth group (BC, CC and AS) and the other examined measures. There are several negative coefficients. One possible explanation could be that the measures focus upon within-category parameters which differ from the network measures. Furthermore, these network measures also do not show strong mutual correlations with each other.

To provide more insight into the associations between the interdisciplinarity measures examined, a cluster solution based on the correlation coefficients is presented in Fig. 2. From the dendrogram, it is clear that these measures cluster in two groups depending on whether or not a dissimilarity matrix is used. Further, among the ones excluding dissimilarity, the measures depending largely on the overlap of SCs (except pro and 1-Pratt) and the network measures are clustered together. Those coming from other fields are also clustered together. This conclusion can also be observed in the heatmap shown in Fig. A1. in the Appendix.

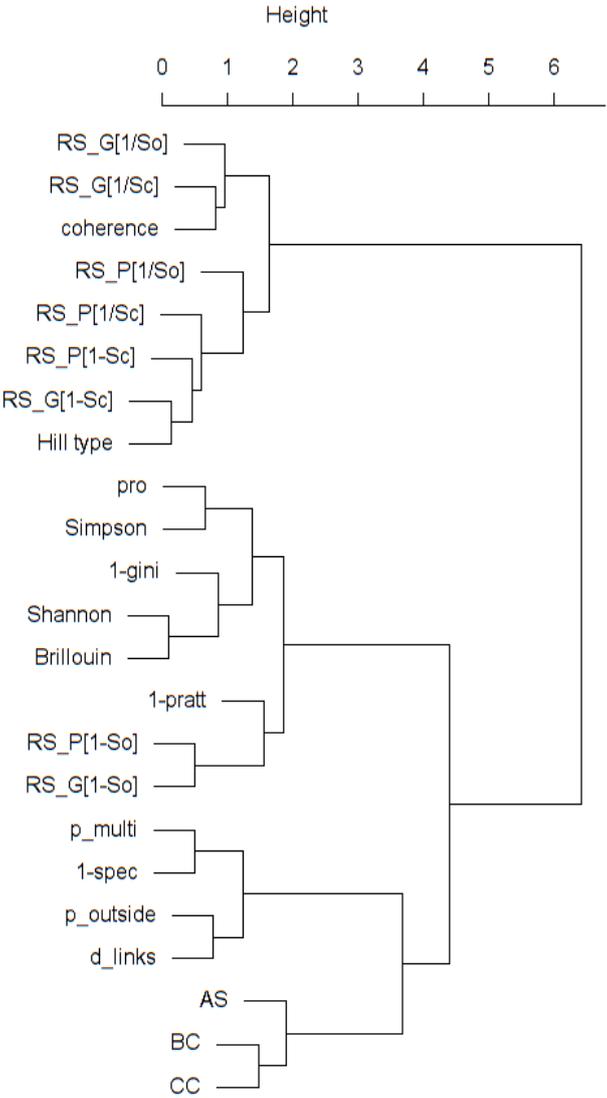

Fig. 2. Cluster dendrogram of interdisciplinarity measures

Table 5. Pearson's correlation coefficients of interdisciplinarity measures

| | 1 | 2 | 3 | 4 | 5 | 6 | 7 | 8 | 9 | 10 | 11 | 12 | 13 | 14 | 15 | 16 | 17 | 18 | 19 | 20 | 21 | 22 | 23 |
|---|---|---|---|---|---|---|---|---|---|---|---|---|---|---|---|---|---|---|---|---|---|---|---|
| 1. p_multi | 1.00 | | | | | | | | | | | | | | | | | | | | | | |
| 2. p_outside | 0.79 | 1.00 | | | | | | | | | | | | | | | | | | | | | |
| 3. pro | 0.44 | 0.45 | 1.00 | | | | | | | | | | | | | | | | | | | | |
| 4. d_links | 0.63 | 0.65 | 0.56 | 1.00 | | | | | | | | | | | | | | | | | | | |
| 5. 1-Pratt | -0.23 | -0.02 | 0.33 | 0.14 | 1.00 | | | | | | | | | | | | | | | | | | |
| 6. 1-Spec | 0.85 | 0.74 | 0.42 | 0.59 | -0.29 | 1.00 | | | | | | | | | | | | | | | | | |
| 7. Simpson | 0.29 | 0.40 | 0.83 | 0.47 | 0.42 | 0.40 | 1.00 | | | | | | | | | | | | | | | | |
| 8. Shannon | 0.19 | 0.37 | 0.64 | 0.43 | 0.55 | 0.33 | 0.86 | 1.00 | | | | | | | | | | | | | | | |
| 9. Brillouin | 0.22 | 0.39 | 0.64 | 0.44 | 0.49 | 0.37 | 0.86 | 1.00 | 1.00 | | | | | | | | | | | | | | |
| 10. 1-Gini | 0.09 | 0.28 | 0.52 | 0.43 | 0.67 | 0.13 | 0.60 | 0.80 | 0.79 | 1.00 | | | | | | | | | | | | | |
| 11. RS_P[1-Sc] | 0.14 | 0.31 | 0.09 | 0.21 | 0.36 | 0.13 | 0.15 | 0.25 | 0.23 | 0.32 | 1.00 | | | | | | | | | | | | |
| 12. RS_G[1-Sc] | 0.13 | 0.27 | 0.00 | 0.16 | 0.35 | 0.09 | 0.01 | 0.17 | 0.15 | 0.32 | 0.91 | 1.00 | | | | | | | | | | | |
| 13. RS_P[1/Sc] | 0.13 | 0.22 | 0.03 | 0.13 | 0.25 | 0.09 | -0.02 | 0.09 | 0.07 | 0.21 | 0.82 | 0.86 | 1.00 | | | | | | | | | | |
| 14. RS_G[1/Sc] | 0.26 | 0.38 | 0.20 | 0.28 | 0.32 | 0.22 | 0.19 | 0.31 | 0.30 | 0.43 | 0.69 | 0.78 | 0.78 | 1.00 | | | | | | | | | |
| 15. RS_P[1-So] | 0.00 | 0.17 | 0.40 | 0.27 | 0.36 | 0.07 | 0.60 | 0.59 | 0.59 | 0.56 | 0.18 | -0.05 | -0.14 | 0.01 | 1.00 | | | | | | | | |
| 16. RS_G[1-So] | 0.03 | 0.22 | 0.39 | 0.28 | 0.43 | 0.08 | 0.60 | 0.65 | 0.64 | 0.68 | 0.29 | 0.15 | 0.00 | 0.22 | 0.93 | 1.00 | | | | | | | |
| 17. RS_P[1/So] | -0.04 | 0.12 | 0.18 | 0.14 | 0.33 | -0.03 | 0.15 | 0.21 | 0.19 | 0.38 | 0.55 | 0.59 | 0.65 | 0.41 | 0.20 | 0.27 | 1.00 | | | | | | |
| 18. RS_G[1/So] | 0.15 | 0.32 | 0.29 | 0.33 | 0.48 | 0.14 | 0.36 | 0.55 | 0.54 | 0.67 | 0.52 | 0.59 | 0.51 | 0.76 | 0.22 | 0.38 | 0.43 | 1.00 | | | | | |
| 19. Hill type | 0.13 | 0.27 | 0.01 | 0.17 | 0.35 | 0.11 | 0.04 | 0.19 | 0.18 | 0.35 | 0.87 | 0.96 | 0.83 | 0.78 | -0.01 | 0.18 | 0.58 | 0.60 | 1.00 | | | | |
| 20. coherence | 0.23 | 0.39 | 0.41 | 0.37 | 0.50 | 0.20 | 0.44 | 0.46 | 0.44 | 0.49 | 0.82 | 0.77 | 0.64 | 0.64 | 0.26 | 0.40 | 0.53 | 0.56 | 0.74 | 1.00 | | | |
| 21. BC | -0.02 | 0.08 | -0.15 | -0.25 | 0.00 | 0.14 | 0.07 | 0.30 | 0.32 | 0.13 | 0.05 | 0.12 | 0.08 | 0.08 | -0.04 | -0.01 | 0.10 | 0.21 | 0.14 | -0.03 | 1.00 | | |
| 22. CC | 0.14 | 0.11 | -0.06 | -0.16 | -0.18 | 0.23 | 0.08 | 0.10 | 0.12 | -0.08 | -0.30 | -0.30 | -0.30 | -0.19 | -0.11 | -0.15 | -0.39 | -0.08 | -0.28 | -0.36 | 0.38 | 1.00 | |
| 23. AS | -0.02 | 0.10 | 0.21 | 0.16 | 0.18 | 0.13 | 0.44 | 0.62 | 0.64 | 0.46 | -0.29 | -0.38 | -0.41 | -0.21 | 0.52 | 0.47 | -0.19 | 0.15 | -0.31 | -0.23 | 0.31 | 0.33 | 1.00 |

*4.2. Distribution of WoS SCs over interdisciplinarity*

Fig. 3 shows the distributions of WoS SCs over interdisciplinarity values. For each histogram in the panel, the x-axis shows the degree of interdisciplinarity and the y-axis shows the frequency of SCs. Some of the histograms reveal considerable differences in the distributions. For example, RS_P[1-$S_o$] and RS_G[1-$S_o$] are left-skewed whereas RS_P[1/$S_o$] and RS_G[1/$S_o$] are highly right-skewed. Moreover, the interdisciplinarity values for some of the measures are concentrated within specific ranges. Taking RS_G[1-$S_o$] as an example, its values are highly concentrated between 0.9 and 1. In addition, it should be noted that some measures are not bounded (e.g. Shannon entropy, Brillouin index, and the Hill-type measure).

One may argue that these distributions are not important because we can transform values into more "suitable" distributions. However, we do believe that they are very useful. First, suppose that we measured interdisciplinarity for a SC using, for instance, RS_G[1-$S_o$] and obtained the value of 0.95. However, it does not seem justified to conclude that this SC is highly interdisciplinary, since RS_G[1-$S_o$] values are almost always close to 1.

Further, it is widely acknowledged that quantitatively determining the validity of an interdisciplinarity measure is challenging, since no benchmarks are available. However, one may for instance expect the distribution of an interdisciplinarity measure to be approximately normal or, alternatively, to be right skewed. Using the distributions presented in Fig. 3, researchers can compare them to these prior expectations. Hence, we believe that studies on interdisciplinarity measures should explicitly state the typical range of interdisciplinarity values and present the empirical distributions.

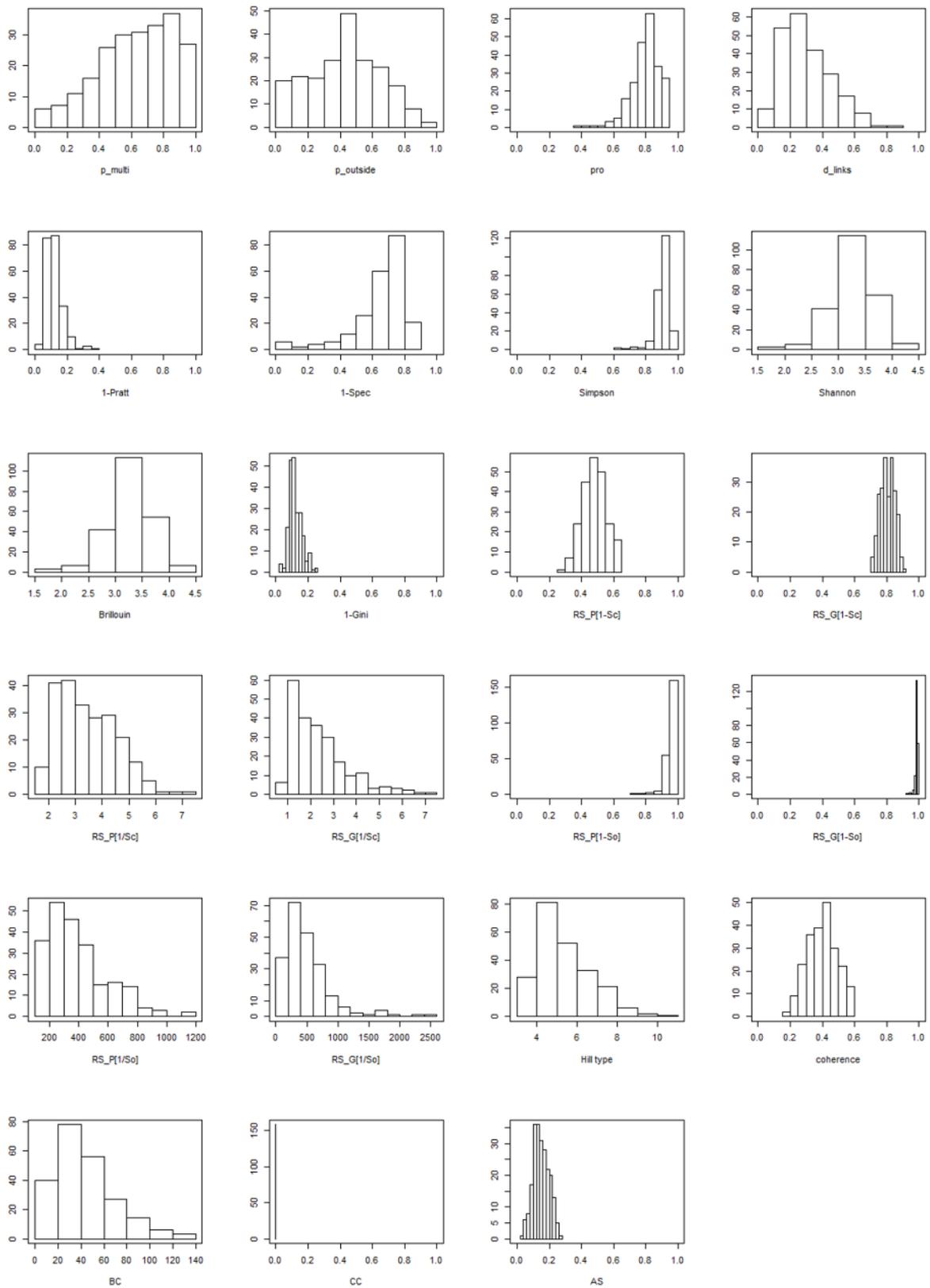

Fig. 3. Distributions of WoS SCs over interdisciplinarity values

*4.3. In-depth analysis of several WoS SCs*

A specific examination of the degree of interdisciplinarity for the WoS SCs based on the various measures provides a more direct impression regarding the effectiveness of these measures. Five WoS SCs are selected; these are NANOSCIENCE & NANOTECHNOLOGY (NANO), BIOCHEMISTRY & MOLECULAR BIOLOGY (BIOM), which previous studies often consider to be highly interdisciplinary (e.g. Aboelela et al., 2006; Porter & Youtie, 2009; Porter et al., 2006), LAW, MATHEMATICS (MATH), which are presumed to show a low degree of interdisciplinarity, and INFORMATION SCIENCE & LIBRARY SCIENCE (LIS). LIS is chosen because most papers investigating interdisciplinarity measures are published in journals attached to this category.

Table 6 presents the different rankings according to the reviewed measures of the five selected SCs among the 224. Instead of reporting the actual interdisciplinarity values obtained from the measures, we provide the ranking obtained after sorting all 224 SCs according to their interdisciplinarity values in decreasing order. Some of the rankings are not in line with our expectations. For instance, NANO is *per se* considered to be an interdisciplinary SC, but its rankings in Table 6 based on the measures that include dissimilarities are generally quite low. Furthermore, measures having mutually strong Spearman's correlations (see Table A2), sometimes lead to conflicting rankings for one specific category. For instance, RS_P[1-$S_c$] and RS_G[1-$S_c$] have a strong Spearman's correlation coefficient (0.91), however, MATH was ranked as 221 and 79 respectively by these measures among 224 SCs. Consequently, given the lack of agreement between the different interdisciplinarity measures, it is very difficult to determine the degree of interdisciplinarity for a particular WoS SC. Measures which are supposed to be similar or reflect similar aspects of IDR can produce very different results.

Further, the distributions in Fig. 3 should also be taken into consideration when we compare the rankings of the five SCs. Some measures have very dense distributions in an extremely short interval. Most rankings based on such measures are therefore not robust as they depend on the third or even fourth decimal and may thus be hard to explain.

Table 6. Interdisciplinarity rankings of the five SCs

| Interdisciplinarity measures | NANO | BIOM | LIS | LAW | MATH |
|---|---:|---:|---:|---:|---:|
| p_multi | 6 | 60 | 185 | 177 | 186 |
| p_outside | 32 | 104 | 140 | 133 | 214 |
| Pro | 21 | 166 | 140 | 213 | 223 |
| d_links | 41 | 165 | 173 | 169 | 221 |
| 1-Pratt | 206 | 106 | 88 | 133 | 224 |
| 1-Spec | 3 | 71 | 137 | 182 | 201 |
| Simpson index | 101 | 112 | 121 | 203 | 223 |
| Shannon entropy | 170 | 74 | 83 | 141 | 224 |
| Brillouin index | 168 | 71 | 81 | 141 | 224 |
| 1-Gini | 201 | 93 | 73 | 97 | 222 |
| RS_P[1-$S_c$] | 192 | 217 | 3 | 80 | 221 |
| RS_G[1-$S_c$] | 203 | 213 | 9 | 34 | 79 |
| RS_P[1/$S_c$] | 181 | 210 | 4 | 39 | 42 |
| RS_G[1/$S_c$] | 175 | 180 | 12 | 52 | 124 |
| RS_P[1-$S_o$] | 189 | 150 | 109 | 100 | 224 |
| RS_G[1-S$_o$] | 207 | 170 | 95 | 39 | 223 |
| RS_P[1/S$_o$] | 203 | 202 | 37 | 4 | 180 |
| RS_G[1/S$_o$] | 197 | 149 | 37 | 43 | 216 |
| Hill-type measure | 203 | 213 | 9 | 34 | 79 |
| Coherence | 209 | 214 | 19 | 152 | 224 |
| BC | 123 | 17 | 30 | 29 | 68 |
| CC | 12 | 3 | 139 | 100 | 105 |
| AS | 88 | 8 | 138 | 89 | 223 |

## 5. Discussion

Based on our analyses, three issues are worth further discussion. First, we discuss the definitions and attributes of IDR, then we focus on interdisciplinarity measures and their operationalization, and finally on their policy implications.

*5.1. Attributes of IDR*

As already indicated, previous studies have argued that the conception of IDR is ambiguous (e.g. Rafols, 2012). We claim the same holds for its definitions in scientometric studies. Based on our review, we found that diversity has been widely seen as the essential and necessary attribute for measuring interdisciplinarity. The attribute of coherence is mainly seen as supplementary and are often ignored in practice. Therefore, most interdisciplinarity measures in scientometric studies use the diversity attribute and most often this attribute alone. For instance, Steele and Stier (2000) state that "[i]n effect, we treat diversity as a proxy measure of interdisciplinarity" (p. 477). This raises the important question of whether diversity in itself is sufficient to capture the concept of interdisciplinarity. We are skeptical.

Generally, we are concerned about the definitions and simplistic indicators used to measure the multidimensional concept of IDR. We especially question to the extent to which diversity is an appropriate attribute that in itself can encompass and reflect the concept. During our review, it became clear that the definition of IDR by the US Committee on Facilitating Interdisciplinary Research and Committee on Science (CFIRCS) was frequently referred to. According to CFIRCS (2005, p. 2):

> "[i]nterdisciplinary research (IDR) is a mode of research by teams or individuals that integrates information, data, techniques, tools, perspectives, concepts, and/or theories from two or more disciplines or bodies of specialized knowledge to advance fundamental understanding or to solve problems whose solutions are beyond the scope of a single discipline or area of research practice".

The definition simply means that IDR would require the integration of knowledge from two or more disciplines. Therefore, we argue that high diversity does not seem to be either a necessary or a sufficient attribute for measuring interdisciplinarity.

In our view, much more work should be done in order to define the concept, identify key attributes, and subsequently empirically examine the construct validity of the measures or composite measures developed to assess interdisciplinarity. Some researchers have indeed argued that the concept of IDR is multidimensional, and hence its attributes should be portrayed using various measures (e.g. Leydesdorff & Rafols 2010; Rafols & Meyer, 2010; Sugimoto &

Weingart, 2014). The recent report from Digital Science states that "no single indicator can unequivocally identify and monitor IDR activity and no present proxy is a demonstrably satisfactory management tool on its own" (p. 9). Our results support these claims inasmuch as we demonstrate that seemingly similar measures produce different results and are sensitive to levels of analysis. This is a serious breach of construct validity.

Few studies have examined theoretical frameworks around interdisciplinarity and linked it to measurement. The main contribution comes from Rafols and colleagues (e.g. Rafols, 2014; Rafols & Meyer, 2010; Rafols et al., 2012). A theoretical framework is needed in order to outline the dimensions of interdisciplinarity and relations between attributes. Unfortunately, such an important endeavor has not received much attention. Instead, "novel" measures, mostly based on diversity, are being proposed continuously. Their relevance, validity and resemblance to other measures are most often overlooked. In this context, we believe that the discussion on which attributes are essential for depicting the nature of interdisciplinarity, as well as the similarity of measures, are necessary and essential.

*5.2. Interdisciplinarity measures and operationalization*

Based on our analyses, we found that even measures with a supposedly similar focus can produce contradictory results when measuring interdisciplinarity using WoS SCs. Inconsistency in our results implies that some measures are problematic for the purpose of describing interdisciplinarity as they do not capture their target attribute. For instance, 1-Pratt and 1-Spec are expected to be consistent. But our results show that this is not the case when applied to WoS SCs. We do not imply that these measures are mathematically wrong, but their validity as measures of interdisciplinarity is doubtful and should be carefully considered.

We also found that the justification for the use of a measure is not always convincing. For example, the Brillouin index is an entropy-based indicator. When it was introduced as an interdisciplinarity measure (e.g. Steele & Stier, 2000; Chang & Huang, 2012; Huang & Chang, 2012), its relation to Shannon entropy was not discussed. Shannon entropy was already used to measure interdisciplinarity, so the supposed merits of the Brillouin index compared to Shannon's entropy should, of course, have been explained in these studies. Our results obtained from Shannon and Brillouin are almost perfectly correlated, which suggests that at least one of them is superfluous. We therefore argue that the introduction and creation of new measures should aim to improve the validity and accuracy of measurement, instead of constantly introducing new and perhaps nearly identical measures. Following the suggestions given by Waltman (2016) on citation impact indicators, we reckon that given the large number of interdisciplinarity measures that already exists, it is not necessary to provide more indicators, especially not indicators relying on the diversity attribute, unless a novel measure has some convincing new merits in relation to validity and accuracy.

The operationalization of interdisciplinarity measures in scientometric studies is relatively chaotic. The report by Digital Science (2016) shows that the degree of interdisciplinarity is influenced by the choice of data sources and classification systems. The present study further demonstrates the tangled and unsustainable situation of measuring interdisciplinarity, with inconsistent outcomes generated by seemingly similar measures.

To be more specific, the present study examines various combinations of RS, demonstrating that they lead to quite different results. Unfortunately, we see that important details have persistently been overlooked in previous studies, for instance, explanations for the choice of cosine formulas (e.g. Porter & Rafols, 2009). Since substantial differences may result from such choices, we suggest that researchers should provide sufficient details on the operationalization of their interdisciplinarity measures and preferably perform sensitivity and robustness analyses.

Also, measures with extremely narrow distributions or without boundary have very little practical use. As shown, interdisciplinarity measures like Shannon entropy and the Hill-type indicator do not have an obvious domain of values. Consequently, it is very difficult to evaluate to what degree a unit is interdisciplinary according to such measures.

*5.3. Interdisciplinarity and policy implication*

The importance of IDR has been widely acknowledged. Many studies argue that it could solve complex problems and promote scientific developments and innovations (Gibbons et al., 1994, see also Rafols et al, 2012; Hollingsworth & Hollingsworth, 2000; Lowe & Phillipson, 2006). As a consequence, "funding agencies in many developed countries are considering enhancing IDR as a topic of priority (Bordons et al. 2004; Rinia, 2007). For instance, research-funding agencies like NSF, Research Councils UK (RCUK), NSFC, and Swedish Research Council (VR) take the promotion of interdisciplinary research an essential task" (Wang, 2016, p. 21).

On the one hand, we observe the enthusiasm of research-funding agencies to encourage and finance IDR. On the other hand, we found that the interdisciplinarity measures in scientometric studies are confusing as they lack validity. The degree of interdisciplinarity for a unit of analysis most likely varies with the choice of measure, data source, and classification system. For a set of units, different measures will most likely produce different rankings between the units. Obviously, this is untenable and of great concern in science policy and research evaluation. It is simply too easy to intentionally influence the outcomes of interdisciplinarity measures. There are too many researcher degrees of freedom (Wicherts et al., 2016).

Interdisciplinarity measures in scientometric studies use publications and citation relations as the data source to identify IDR. In other words, we understand interdisciplinarity from a bibliometric perspective. However, we are indifferent to how other stakeholders, like policy makers, understand IDR. Interdisciplinarity measures in scientometric studies may be able to deal with some aspects of interdisciplinarity but not others. Hence, it is important and necessary to thoroughly state which aspects (attributes) of interdisciplinarity they actually depict when reporting studies of interdisciplinarity. In addition, it is also important not to be blinded by measures relying on bibliometric methods. They tend to produce a "tunnel vision" where this is the only way to measure interdisciplinarity.

## 6. Conclusions

The present article aims to systematically examine the consistency and relation between interdisciplinarity measures based on bibliometric methods. We first examined these measures focusing on the WoS SCs. Based on correlations and clustering, we found that the 23 reviewed measures can be roughly classified into two groups depending on whether or not a dissimilarity matrix is used. Measures in the same cluster tend to have fairly strong mutual correlations, but are weakly correlated with the measures in other groups. However, while some measures are supposed to measure similar aspects, they nevertheless turn out to be inconsistent (e.g. 1-Pratt and 1-Spec). Further, highly correlated measures also provide very conflicting results when they are used to measure interdisciplinarity of WoS SCs (e.g. RS_P[1-Sc] and RS_G[1-Sc]). In other words, a high overall correlation does not equal homogenous results. Even though we see strong correlations between some measures at the aggregate level, substantially different ranking are observed in a few cases at the individual SC-level. The reason for this is unclear. Finally, histograms showing the distribution of interdisciplinarity values over different WoS SCs reveal tight distributions for some measures as their values are concentrated in limited intervals (e.g. CC and AS). These measures may be problematic when used in practice. We therefore conclude that the degree of interdisciplinarity for a unit of analysis is strongly dependent on the choice of measures.

The findings in our study complement the conclusions in the report from Digital Science (2016): "choice of data, methodology and indicators can produce seriously inconsistent results despite a common set of disciplines and countries" (p. 2). Our results further demonstrate that inconsistent and even conflicting findings can come out of analyses based on the same data source and the same classification. The current state of interdisciplinarity measurement is confusing and unsustainable. Interdisciplinarity is a multidimensional concept and measures should reflect these dimensions through various different attributes either as single or composite indicators. However, we find that the definitions of interdisciplinarity are quite similar and hardly multidimensional. This fact makes it even more complicated to interpret the inconsistent values we obtained from these presumably similar measures.

The validity and robustness of interdisciplinarity studies using bibliometric methods should be questioned. As it is, measures and their values are inconsistent and non-robust. This can lead to an untenable situation where the choice of (arbitrary) measures determines the degree of interdisciplinarity, but not the underlying nature of research which they are supposed to characterize. We therefore suggest that future studies on interdisciplinarity focus more upon the theoretical and measurement frameworks, and put more effort into examining the validity and relations between the definition and the use of measures. In addition, we recommend that we simply stop using the current interdisciplinarity measures in policy studies, as they have no warrant.

**Acknowledgment**

The authors would like to thank Ismael Rafols, Ludo Waltman, Gaël Dubus, and reviewers for their valuable comments and suggestions.

**Appendix**

Table A1. Definitions of IDR from a bibliometric perspective

| Study | Definition |
|---|---|
| Steele & Stier (2000) | "In effect, we treat diversity as a proxy measure of interdisciplinarity. Theoretical support for this approach is provided by Gibbons, Limoges, Nowotny, Schwartzman, Scott, & Trow (1994), who asserted that, by definition, interdisciplinarity involves heterogeneity, specifically a diversity of individuals, skills, experiences, institutions, linkages, and locations." (p. 477) |
| Morillo et al. (2001) | "Strictly speaking, we consider "multidisciplinarity" as a basic situation in which elements from different disciplines are present, whilst "interdisciplinarity" is a more advanced stage of the relationship between disciplines in which integration between them is attained." (p.204) |
| Morillo et al. (2003) | "Interdisciplinary research leads to the creation of a theoretical, conceptual, and methodological identity, so more coherent and integrated results are obtained." (p. 1237) |
| Leydesdorff (2007) | "… it occurred to me that the interdisciplinarity of journals corresponds with their visible position in the vector space" (p. 1305) |
| Porter, et al. (2007); Porter, et al. (2008); Wang (2016); Zhang, et al. (2016) | "We apply the following definition, based on a National Academies report: Interdisciplinary research (IDR) is a mode of research by teams or individuals that integrates<br>• perspectives/concepts/theories and/or<br>• tools/techniques and/or<br>• information/data<br>from two or more bodies of specialized knowledge or research practice." (Porter, et al., 2007, p. 119) |
| Porter & Meyer (2009) | "Thus, the process of integrating different bodies of knowledge rather than transgression of disciplinary boundaries per se, has been identified as the key aspect of so-called 'interdisciplinary research.' (National Academies 2005)." (p. 264) |
| Porter & Rafols (2009) | "This report operationally defined interdisciplinary research as:<br>a mode of research by teams or individuals that integrates perspectives/concepts/theories and/or<br>tools/techniques and/or<br>information/data<br>from two or more bodies of knowledge or research practice." (p. 720)<br>"Understood as knowledge integration, interdisciplinarity is not the opposite of specialization. … Our investigation here does not concern the degree of topic specialization of research but the degree that it relies on distinct." (p. 720) |
| Leydesdorff & Rafols (2011) | "Furthermore, interdisciplinarity may be a transient phenomenon. As a new specialty emerges, it may draw heavily on its mother disciplines/specialties, but as it matures a set of potentially new journals can be expected to cite one another increasingly, and thus to develop a type of closure that is typical of "disciplinarity" (Van den Besselaar & Leydesdorff, 1996). Interdisciplinarity, however, may mean something different at the top of the journal hierarchy (as in the case of Science and Nature) than at the bottom, where one has to draw on different bodies of knowledge for the sake of the application (e.g., in engineering). Similarly, in the clinic one may be more inclined to integrate knowledge from different specialties at the bedside than a laboratory where the focus is on specialization and refinement." (p. 88) |
| Chang & Huang (2012) | "A common feature of interdisciplinarity as it is manifested in a variety of research activities is the transfer of information across disciplines (Porter, Roessner, Cohen, & Perreault, 2006). Pierce (1999) grouped interdisciplinary information transfer into three types: the citation of references from different disciplines, the co-authoring of articles by researchers from |
| | different disciplines, and the publishing of works within other disciplines. Such transfer implies that the degree of interdisciplinarity of a specific discipline can be determined by analyzing the discipline distribution of references and co-authors in publications." (p. 22) |
| Huang & Chang (2012) | "The concept of interdisciplinarity has been discussed by many researchers (Huutoniemi et al. 2010; Leydesdorff and Probst 2009; Rosenfield 1992; Tijssen 1992), and can be defined as the use of knowledge, methods, techniques, and devices as a result of scientific activities from other fields (Tijssen 1992)." (p. 790) |
| Rafols, et al. (2012) | "We propose to investigate interdisciplinarity from two perspectives, each of which we claim has more general applicability. The first is by means of the widely used conceptualisation of interdisciplinarity as knowledge integration (National Academies, 2004; Porter et al., 2006), which is perceived as crucial for innovation or solving social problems. The second is by conceptualising interdisciplinarity as a form of research that lies outside or in between established practices, i.e. in terms of intermediation (Leydesdorff, 2007a)." (p.1265) |
| Silva et al. (2013). | "In a sense, this movement brought science closer to the paradigm adopted by Greek philosophers who treated Nature as a landscape of knowledge glued together in an indivisible discipline. Not surprisingly, in recent years new areas have been established with this interdisciplinary character, as is the case of nanoscience and nanotechnology, in addition to new disciplines arising from the merging of two or more areas, such as computational biology and biomolecular physics." (p. 469) |
| Soo & Kampis (2012). | "… diversity measures are clearly associated with the degree multidisciplinarity. On the other hand, the notion of interdisciplinary research (IDR) is decomposed into two differing perspectives: on one account, IDR is conceived as knowledge integration, an indicator of which is the degree of overall interrelatedness of the units of analysis constituting a portfolio." (p. 871).<br>"On the other account, however, interdisciplinarity is viewed as intermediation between knowledge domains, embodied in publication sets positioned between more established clusters of journals or fields." (p. 871) |
| Wang et al. (2015). | "For interdisciplinary research, integrating knowledge from more disciplines contributes to potential more broadly useful outcomes." (p. 11) |
| Rodriguez (2017). | "Finally, interdisciplinarity entails the integration of 'separate disciplinary data, methods, tools, concepts and theories in order to create a holistic view or common understanding of a complex issue, question or problem'." (p. 619) |

Table A2. Spearman's correlation coefficients of interdisciplinarity measures

|  | 1 | 2 | 3 | 4 | 5 | 6 | 7 | 8 | 9 | 10 | 11 | 12 | 13 | 14 | 15 | 16 | 17 | 18 | 19 | 20 | 21 | 22 | 23 |
|---|---|---|---|---|---|---|---|---|---|---|---|---|---|---|---|---|---|---|---|---|---|---|---|
| 1. p_multi | 1.00 | | | | | | | | | | | | | | | | | | | | | | |
| 2. p_outside | 0.78 | 1.00 | | | | | | | | | | | | | | | | | | | | | |
| 3. pro | 0.45 | 0.44 | 1.00 | | | | | | | | | | | | | | | | | | | | |
| 04. d_links | 0.64 | 0.66 | 0.59 | 1.00 | | | | | | | | | | | | | | | | | | | |
| 5. 1-Pratt | -0.07 | 0.14 | 0.44 | 0.27 | 1.00 | | | | | | | | | | | | | | | | | | |
| 6. 1-Spec | 0.82 | 0.78 | 0.47 | 0.63 | 0.05 | 1.00 | | | | | | | | | | | | | | | | | |
| 7. Simpson | 0.26 | 0.40 | 0.78 | 0.54 | 0.69 | 0.43 | 1.00 | | | | | | | | | | | | | | | | |
| 8. Shannon | 0.17 | 0.35 | 0.59 | 0.43 | 0.78 | 0.36 | 0.92 | 1.00 | | | | | | | | | | | | | | | |
| 9. Brillouin | 0.19 | 0.36 | 0.59 | 0.44 | 0.75 | 0.38 | 0.92 | 1.00 | 1.00 | | | | | | | | | | | | | | |
| 10. 1-Gini | 0.09 | 0.27 | 0.54 | 0.40 | 0.85 | 0.20 | 0.77 | 0.84 | 0.83 | 1.00 | | | | | | | | | | | | | |
| 11. RS_P[1-Sc] | 0.15 | 0.31 | 0.09 | 0.21 | 0.41 | 0.18 | 0.18 | 0.26 | 0.24 | 0.37 | 1.00 | | | | | | | | | | | | |
| 12. RS_G[1-Sc] | 0.16 | 0.27 | 0.04 | 0.14 | 0.37 | 0.17 | 0.11 | 0.21 | 0.19 | 0.33 | 0.91 | 1.00 | | | | | | | | | | | |
| 13. RS_P[1/Sc] | 0.15 | 0.22 | 0.08 | 0.11 | 0.28 | 0.15 | 0.07 | 0.13 | 0.12 | 0.25 | 0.83 | 0.90 | 1.00 | | | | | | | | | | |
| 14. RS_G[1/Sc] | 0.26 | 0.35 | 0.24 | 0.25 | 0.42 | 0.31 | 0.29 | 0.34 | 0.33 | 0.47 | 0.77 | 0.87 | 0.88 | 1.00 | | | | | | | | | |
| 15. RS_P[1-So] | -0.10 | 0.08 | 0.28 | 0.30 | 0.56 | -0.02 | 0.51 | 0.54 | 0.53 | 0.68 | 0.13 | -0.03 | -0.11 | -0.05 | 1.00 | | | | | | | | |
| 16. RS_G[1-So] | -0.02 | 0.18 | 0.35 | 0.36 | 0.73 | 0.09 | 0.62 | 0.69 | 0.68 | 0.84 | 0.41 | 0.33 | 0.19 | 0.31 | 0.87 | 1.00 | | | | | | | |
| 17. RS_P[1/So] | -0.02 | 0.13 | 0.24 | 0.15 | 0.43 | -0.03 | 0.23 | 0.25 | 0.24 | 0.42 | 0.65 | 0.65 | 0.75 | 0.62 | 0.28 | 0.42 | 1.00 | | | | | | |
| 18. RS_G[1/So] | 0.16 | 0.34 | 0.36 | 0.29 | 0.62 | 0.28 | 0.51 | 0.58 | 0.58 | 0.68 | 0.66 | 0.71 | 0.68 | 0.84 | 0.25 | 0.56 | 0.67 | 1.00 | | | | | |
| 19. Hill type | 0.16 | 0.27 | 0.04 | 0.14 | 0.37 | 0.17 | 0.11 | 0.21 | 0.19 | 0.33 | 0.91 | 1.00 | 0.90 | 0.87 | -0.03 | 0.33 | 0.65 | 0.71 | 1.00 | | | | |
| 20. coherence | 0.25 | 0.39 | 0.38 | 0.36 | 0.55 | 0.25 | 0.42 | 0.44 | 0.42 | 0.49 | 0.81 | 0.77 | 0.66 | 0.69 | 0.19 | 0.47 | 0.60 | 0.67 | 0.77 | 1.00 | | | |
| 21. BC | 0.00 | 0.11 | -0.17 | -0.26 | 0.06 | 0.14 | 0.10 | 0.30 | 0.32 | 0.10 | 0.03 | 0.09 | 0.06 | 0.10 | -0.09 | -0.02 | -0.03 | 0.19 | 0.09 | -0.06 | 1.00 | | |
| 22. CC | 0.21 | 0.21 | -0.10 | -0.04 | -0.28 | 0.38 | 0.09 | 0.14 | 0.17 | -0.13 | -0.30 | -0.30 | -0.33 | -0.19 | -0.25 | -0.26 | -0.53 | -0.13 | -0.30 | -0.39 | 0.55 | 1.00 | |
| 23. AS | -0.04 | 0.09 | 0.19 | 0.17 | 0.31 | 0.13 | 0.49 | 0.58 | 0.59 | 0.44 | -0.29 | -0.38 | -0.47 | -0.31 | 0.58 | 0.46 | -0.22 | 0.02 | -0.38 | -0.24 | 0.33 | 0.40 | 1.00 |

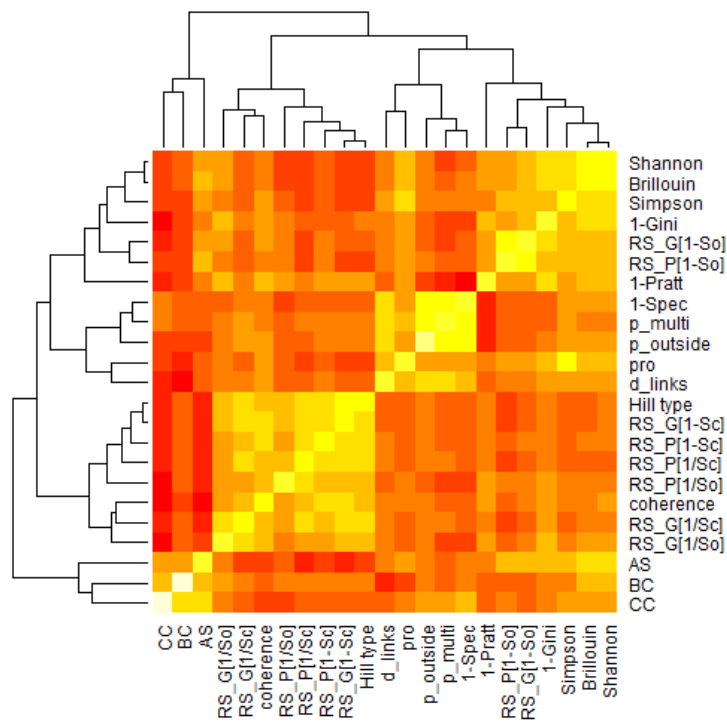

Fig. A1. Heatmap of interdisciplinarity measures

Shannon, C. E. (2001). A mathematical theory of communication. *ACM SIGMOBILE Mobile Computing and Communications Review*, 5(1), 3-55.

Silva, F. N., Rodrigues, F. A., Oliveira, O. N., & Costa, L. D. F. (2013). Quantifying the interdisciplinarity of scientific journals and fields. *Journal of Informetrics*, 7(2), 469-477.

Simpson, E. H. (1949). Measurement of diversity. *Nature*, 163, 688.

Soós, S., & Kampis, G. (2012). Beyond the basemap of science: mapping multiple structures in research portfolios: Evidence from Hungary. *Scientometrics*, 93(3), 869-891.

Steele, T. W., & Stier, J. C. (2000). The impact of interdisciplinary research in the environmental sciences: A forestry case study. *Journal of the American Society for Information Science*, 51(5), 476-484.

Stirling, A. (2007). A general framework for analysing diversity in science, technology and society. *Journal of the Royal Society Interface*, 4(15), 707-719.

Sugimoto, C. R., & Weingart, S. (2015). The kaleidoscope of disciplinarity. *Journal of Documentation*, 71(4), 775-794.

Wang, J., Thijs, B., & Glänzel, W. (2015). Interdisciplinarity and impact: Distinct effects of variety, balance, and disparity. *PloS one*, 10(5), e0127298.

Wicherts, J. M., Veldkamp, C. L. S., Augusteijn, H. E. M., Bakker, M., van Aert, R. C. M., & van Assen, M. A. L. M. (2016). Degrees of freedom in planning, running, analyzing, and reporting psychological studies: A checklist to avoid p-hacking. *Frontiers in Psychology*, 7, 1832.

Whitley, R. (2000). *The Intellectual and Social Organization of the Sciences.* Oxford: Oxford University Press.

Zhang, L., Rousseau, R., & Glänzel, W. (2016). Diversity of references as an indicator for interdisciplinarity of journals: Taking similarity between subject fields into account. *Journal of the American Society for Information Science and Technology*, 67(5), 1257-1265.

Zhou, Q.J., Rousseau, R., Yang, L.Y., Yue, T., & Yang, G.L. (2012). A general framework for describing diversity within systems and similarity between systems with applications in informetrics. *Scientometrics*, 93(3), 787–812.27

**Author contributions**

Qi Wang: conceptualization, data curation, formal analysis, investigation, methodology, software, validation, visualization, writing – original draft

Jesper Wiborg Schneider: conceptualization, methodology, resources, supervision, writing – review & editing

**Competing interests**

The authors have no competing interests.

**Data availability**

The raw bibliometric data were obtained from Clarivate Analytics' Web of Science database. A Web of Science license is required to access the data. The values of the different interdisciplinarity measures at the level of Web of Science subject categories and the table of research areas used to measure p_outside are available at
https://kth.box.com/s/zbcfmvpuhmhwvl8ql1u522snyht34y50.

**Funding**

No funding has been received.